# Analysis of Distributed ADMM Algorithm for Consensus Optimization in Presence of Error

Layla Majzoobi+, Farshad Lahouti*
+School of Electrical and Computer Engineering, University of Tehran
*Electrical Engineering Department, California Institute of Technology

**Abstract**

ADMM is a popular algorithm for solving convex optimization problems. Applying this algorithm to distributed consensus optimization problem results in a fully distributed iterative solution which relies on processing at the nodes and communication between neighbors. Local computations usually suffer from different types of errors, due to e.g., observation or quantization noise, which can degrade the performance of the algorithm. In this work, we focus on analyzing the convergence behavior of distributed ADMM for consensus optimization in presence of additive node error. We specifically show that (a noisy) ADMM converges linearly under certain conditions and also examine the associated convergence point. Numerical results are provided which demonstrate the effectiveness of the presented analysis.

## I. INTRODUCTION

The advent of big data, the Internet of Things and cyber-physical systems often calls for optimization algorithms that run on distributed data for learning, inference, action and control. Many interesting problems in this domain are cast as convex optimization problems; examples include network flow control, feature extraction and (power system) state estimation. Considerable research has been devoted to distributed optimization methods [1]- [7]. These algorithms may be characterized by their range of applicability, convergence behavior and performance, and computational complexity. Alternating Direction Method of Multipliers (ADMM) is an iterative optimization algorithm which can be implemented in a distributed manner [8]. Applicability to a wide range of problems and convergence to a modest accuracy within a few tens of iterations make ADMM very useful in practice. As a critical characteristic of an iterative algorithm, the convergence behavior of ADMM is analyzed in [9]- [11].

Consensus optimization is a popular distributed optimization problem which arises in various domains [3], [12], [13] and is formulated by

$$\min_{\tilde{\mathbf{x}}} \quad \sum_{i=1}^{N} f_i(\tilde{\mathbf{x}}) \tag{1}$$

in which $\tilde{\mathbf{x}} \in \mathbb{R}^n$ and $f_i(\tilde{\mathbf{x}}) : \mathbb{R}^n \to \mathbb{R}$. The goal is to optimize the sum of local objective functions, $f_i(\tilde{\mathbf{x}}), i \in \{1, ..., N\}$, over a global variable $\tilde{\mathbf{x}}$ at $N$ agents. ADMM-based algorithms for consensus optimization problems are proposed in [14]- [16]. The convergence rate of ADMM in this setting and the corresponding effects of different parameters such as network topology are studied in [16].

The local objective functions in a consensus are usually computed based on some observations and measurements, which are noisy in general. The computed results at a given node are always quantized prior to communication to another node in an iteration. In addition, certain local computations are too complex to carry out exactly and are usually replaced by approximations. Some inexact variants of ADMM algorithm are proposed in [14], [17]. These errors, which we refer to as computation errors or noise, are very common in practical settings, and could substantially affect the performance of distributed optimization methods.

In this work, we focus on analyzing the convergence behavior of distributed ADMM for consensus optimization in presence of computation error. We specifically show that (a noisy) ADMM converges linearly under certain conditions and also examine the associated convergence point. Numerical results are provided which demonstrate the effectiveness of the presented analysis and the effects of certain parameters on performance.

## II. PRELIMINARIES

Consider a network of $N$ nodes that intend to solve the optimization problem (1) in a distributed manner and over a given network with $E$ undirected links. The network can be represented by the graph $\mathcal{G} = (\mathcal{V}, \mathcal{A})$ where, $\mathcal{V}$ is set of nodes, $|\mathcal{V}| = N$, and $\mathcal{A}$ is set of arcs, $|\mathcal{A}| = 2E$. Communications of nodes are synchronous and restricted to the neighbors.

**Algorithm 1**

Input functions $f_i$; Initialization: for all $i \in \mathcal{V}$, set $\mathbf{x}_i^0 = \alpha_i^0 = 0_{n \times 1} \in \mathbb{R}^n$ ; Set algorithm parameter $c > 0$; $k = 0$:

1: **for all** $k = 0, 1, \ldots$, every node $i$ **do**
2:   Update $\mathbf{x}_i^{k+1}$ by solving $\nabla f_i(\mathbf{x}_i^{k+1}) + \alpha_i^k + 2c|\mathcal{N}_i|\mathbf{x}_i^{k+1} - c(|\mathcal{N}_i|\mathbf{x}_i^k + \sum_{j \in \mathcal{N}_i} \mathbf{x}_j^k) = \mathbf{0}$ ;
3:   Update $\alpha_i^{k+1} = \alpha_i^k + c(|\mathcal{N}_i|\mathbf{x}_i^{k+1} - \sum_{j \in \mathcal{N}_i} \mathbf{x}_j^{k+1})$;

In order to apply ADMM algorithm to problem (1), it is reformulated as [16]

$$\min_{\mathbf{x}_i, \mathbf{z}_{ij}} \sum_{i=1}^N f_i(\mathbf{x}_i) \qquad (2)$$
$$s.t. \quad \mathbf{x}_i = \mathbf{z}_{ij}, \quad \mathbf{x}_j = \mathbf{z}_{ij} \quad \forall (i,j) \in \mathcal{A}$$

where $\mathbf{x}_i$ is the copy of the global optimization variable $\tilde{\mathbf{x}}$ at node $i$ and $\mathbf{z}_{ij}$ is the variable which enforces equality of $\mathbf{x}_i$ and $\mathbf{x}_j$ at the optimized point. Concatenating $\mathbf{x}_i$'s and $\mathbf{z}_{ij}$'s respectively into $\mathbf{x} \in \mathbb{R}^{nN}$ and $\mathbf{z} \in \mathbb{R}^{2En}$, (2) can be rewritten in a matrix form as follows

$$\min_{\mathbf{x}, \mathbf{z}} f(\mathbf{x}) + g(\mathbf{z}); \quad s.t. \quad \mathbf{Ax} + \mathbf{Bz} = 0 \qquad (3)$$

where $f(\mathbf{x}) = \sum_{i=1}^N f_i(\mathbf{x}_i)$, $g(\mathbf{z}) = 0$, $\mathbf{A} = [\mathbf{A}_1; \mathbf{A}_2]; \mathbf{A}_1, \mathbf{A}_2 \in \mathbb{R}^{2En \times Nn}$ are both composed of $2E \times N$ blocks of $n \times n$ matrices. If $(i,j) \in \mathcal{A}$ and $\mathbf{z}_{ij}$ is the $q$'th block of $\mathbf{z}$, then the $(q,i)$ block of $\mathbf{A}_1$ and the $(q,j)$ block of $\mathbf{A}_2$ are $n \times n$ identity matrices $\mathbf{I}_n$; otherwise the corresponding blocks are $n \times n$ zero matrices $\mathbf{0}_n$. Also, we have $\mathbf{B} = [-\mathbf{I}_{2En}; -\mathbf{I}_{2En}]$. Forming augmented Lagrangian and applying ADMM algorithm we have

$$\nabla f(\mathbf{x}^{k+1}) + \mathbf{A}^T \boldsymbol{\lambda}^k + c\mathbf{A}^T(\mathbf{Ax}^{k+1} + \mathbf{Bz}^k) = \mathbf{0} \qquad (4a)$$
$$\mathbf{B}^T \boldsymbol{\lambda}^k + c\mathbf{B}^T(\mathbf{Ax}^{k+1} + \mathbf{Bz}^{k+1}) = \mathbf{0} \qquad (4b)$$
$$\boldsymbol{\lambda}^{k+1} - \boldsymbol{\lambda}^k - c(\mathbf{Ax}^{k+1} + \mathbf{Bz}^{k+1}) = \mathbf{0} \qquad (4c)$$

where $\boldsymbol{\lambda}$ and $c$ are Lagrange multiplier and ADMM parameter, respectively; $\nabla f(\mathbf{x}^{k+1})$ is gradient or subgradient of $f(\mathbf{x})$ at $\mathbf{x}^{k+1}$ depending on the differentiability of $f$. Considering $\boldsymbol{\lambda} = [\boldsymbol{\beta}; \boldsymbol{\gamma}]$ with $\boldsymbol{\beta}, \boldsymbol{\gamma} \in \mathbb{R}^{2En}$, $\mathbf{M}_+ = \mathbf{A}_1^T + \mathbf{A}_2^T$ and $\mathbf{M}_- = \mathbf{A}_1^T - \mathbf{A}_2^T$, it cab be shown that $\boldsymbol{\beta} = -\boldsymbol{\gamma}$ and with some algebraic manipulations [16] we have

$$\nabla f(\mathbf{x}^{k+1}) + \mathbf{M}_- \boldsymbol{\beta}^{k+1} - c\mathbf{M}_+(\mathbf{z}^k - \mathbf{z}^{k+1}) = \mathbf{0} \qquad (5a)$$
$$\boldsymbol{\beta}^{k+1} - \boldsymbol{\beta}^k - \frac{c}{2}\mathbf{M}_-^T \mathbf{x}^{k+1} = \mathbf{0} \qquad (5b)$$
$$\frac{1}{2}\mathbf{M}_+^T \mathbf{x}^k - \mathbf{z}^k = \mathbf{0} \qquad (5c)$$

Defining $\boldsymbol{\alpha} = [\boldsymbol{\alpha}_1; \boldsymbol{\alpha}_2; \ldots; \boldsymbol{\alpha}_N] = \mathbf{M}_- \boldsymbol{\beta} \in \mathbb{R}^{Nn}$, a relatively simple fully decentralized algorithm is obtained in [16] and presented in Algorithm 1. In this Algorithm, $\mathcal{N}_i$ is the set of neighbors of node $i$.

As stated, we assume that the neighboring nodes only exchange a noisy version of their messages in step 2 of the Algorithm 1. We assume this so-called computation error $\mathbf{e}_{x,i}^k$, that is for example due to quantization of $x_i^k$ at node $i$ in iteration $k$, is additive and independent, identically distributed. We have

$$\hat{\mathbf{x}}_i^k = \mathbf{x}_i^k + \mathbf{e}_{x,i}^k \qquad \hat{\mathbf{x}} = \mathbf{x} + \mathbf{e}_x^k \qquad (6)$$

where $\hat{\mathbf{x}}_i^k$ is the message that node $i \in \mathcal{N}$ communicates to its neighbours at iteration $k$; and $\hat{\mathbf{x}} \in \mathbb{R}^{nN}$ and $\mathbf{e}_x^k \in \mathbb{R}^{nN}$ denote the concatenated versions of $\hat{\mathbf{x}}_i^k$ and $\mathbf{e}_{x,i}^k, \forall i$. In the following Section, we analyze the convergence behavior of Algorithm 1 while taking into account the computation error.

## III. Convergence Analysis of Noisy ADMM

Our convergence analysis is based on (5). Replacing $\mathbf{x}^k$ in (5c), with $\hat{\mathbf{x}}^k$ from (6), we have

$$\hat{\mathbf{z}}^k = \frac{1}{2}\mathbf{M}_+^T \mathbf{x}^k + \frac{1}{2}\mathbf{M}_+^T \mathbf{e}_x^k = \mathbf{z}^k + \mathbf{e}_z^k \qquad (7)$$

It means that because of computation error at iteration $k$, $\mathbf{z}^k$ is now $\mathbf{z}^k + \mathbf{e}_z^k$. Hence (5) can be rewritten as follows

$$\nabla f(\mathbf{x}^{k+1}) + \mathbf{M}_- \boldsymbol{\beta}^{k+1} - c\mathbf{M}_+(\mathbf{z}^k + \mathbf{e}_z^k - \mathbf{z}^{k+1}) = 0 \qquad (8a)$$
$$\boldsymbol{\beta}^{k+1} - \boldsymbol{\beta}^k - \frac{c}{2}\mathbf{M}_-^T \mathbf{x}^{k+1} = 0 \qquad (8b)$$

$$\frac{1}{2}\mathbf{M}_+^T \mathbf{x}^k - \mathbf{z}^k = 0 \tag{8c}$$

Let $\mathbf{u}^k \triangleq \begin{pmatrix} \mathbf{z}^k \\ \boldsymbol{\beta}^k \end{pmatrix}$, $\mathbf{u}^* \triangleq \begin{pmatrix} \mathbf{z}^* \\ \boldsymbol{\beta}^* \end{pmatrix}$, $\mathbf{G} \triangleq \begin{pmatrix} c\mathbf{I}_{2En} & 0_{2En} \\ 0_{2En} & \frac{1}{c}\mathbf{I}_{2En} \end{pmatrix}$. We next analyze the convergence behavior of $\|\mathbf{u} - \mathbf{u}^*\|_G$ and show that its Q-linear convergence results in R-linear convergence of $\|\mathbf{x}^k - \mathbf{x}^*\|_2$.

**Theorem 1.** Consider optimization problem (3) and iterative solution (8) for optimal points $\mathbf{x}^*$ and $\mathbf{z}^*$. Assume that $f_i$'s ($f$) are strongly convex functions with moduli $m_{f_i}$ ($m_f$) and have Lipschitz continuous gradients $\nabla f_i$ ($\nabla f$) with constant $M_{f_i}$ ($M_f$). Suppose that initial value $\boldsymbol{\beta}^0$ lies in the column space of $\mathbf{M}_-^T$. If $\|\mathbf{e}_z^k\|_2 \leq \|\mathbf{x}^{k+1} - \mathbf{x}^*\|_2$ and $m_f - \frac{c}{2}\sigma_{max}^2(\mathbf{M}_+) \geq 0$, then $\forall \mu > 1$

$$\|\mathbf{u}^{k+1} - \mathbf{u}^*\|_G^2 \leq \frac{1}{1+\delta}\|\mathbf{u}^k - \mathbf{u}^*\|_G^2 \tag{9}$$

and

$$\|\mathbf{x}^{k+1} - \mathbf{x}^*\|_2^2 \leq \frac{1}{m_f - \frac{c}{2}\sigma_{max}^2(\mathbf{M}_+)}\|\mathbf{u}^k - \mathbf{u}^*\|_G^2 \tag{10}$$

where

$$\delta = \min\left\{\frac{m_f - \frac{c\sigma_{max}(\mathbf{M}_+)}{2}}{a}, \frac{1}{b}\right\}$$

$$a = \frac{c}{4}\sigma_{max}^2(\mathbf{M}_+) + \frac{2\mu M_f^2}{c\tilde{\sigma}_{min}^2(\mathbf{M}_-)} + \frac{4c\sigma_{max}^2(\mathbf{M}_+)\sigma_{max}^2(\mathbf{M}_-)}{\tilde{\sigma}_{min}^4(\mathbf{M}_-)}$$

$$b = \frac{2\sigma_{max}^2(\mathbf{M}_+)}{(1-\frac{1}{\mu})\tilde{\sigma}_{min}^2(\mathbf{M}_-)}$$

and $\sigma_{max}(\mathbf{D})$ and $\tilde{\sigma}_{min}(\mathbf{D})$ are maximum singular value and minimum non-zero singular value of $\mathbf{D}$.

*Proof.* The optimal value calculated using (5) satisfies KKT conditions, we have

$$\nabla f(\mathbf{x}^*) + \mathbf{M}_- \boldsymbol{\beta}^* = 0; \quad \mathbf{M}_-^T \mathbf{x}^* = 0; \quad \frac{1}{2}\mathbf{M}_+^T \mathbf{x}^* - \mathbf{z}^* = 0 \tag{11}$$

Combining (8) and (11), we have

$$\nabla f(\mathbf{x}^{k+1}) - \nabla f(\mathbf{x}^*) = c\mathbf{M}_+(\mathbf{z}^k + \mathbf{e}_z^k - \mathbf{z}^{k+1}) - \mathbf{M}_-(\boldsymbol{\beta}^{k+1} - \boldsymbol{\beta}^*) \tag{12a}$$

$$\frac{c}{2}\mathbf{M}_-^T(\mathbf{x}^{k+1} - \mathbf{x}^*) = \boldsymbol{\beta}^{k+1} - \boldsymbol{\beta}^k \tag{12b}$$

$$\frac{1}{2}\mathbf{M}_+^T(\mathbf{x}^{k+1} - \mathbf{x}^*) = \mathbf{z}^{k+1} - \mathbf{z}^* \tag{12c}$$

Based on strong convexity assumption, we have

$$m_f\|\mathbf{x}^{k+1} - \mathbf{x}^*\|_2^2 \leq \langle \mathbf{x}^{k+1} - \mathbf{x}^*, \nabla f(\mathbf{x}^{k+1}) - \nabla f(\mathbf{x}^*) \rangle \tag{13}$$

where $\langle \mathbf{v}, \mathbf{x} \rangle$ is the inner product of $\mathbf{v}$ and $\mathbf{w}$. Substituting (12a) in RHS of (13) and using (12b) and (12c), we obtain

$$\langle \mathbf{x}^{k+1} - \mathbf{x}^*, \nabla f(\mathbf{x}^{k+1}) - \nabla f(\mathbf{x}^*) \rangle$$
$$= 2c\langle \mathbf{z}^{k+1} - \mathbf{z}^*, \mathbf{z}^k - \mathbf{z}^{k+1} \rangle + 2c\langle \mathbf{z}^{k+1} - \mathbf{z}^*, \mathbf{e}_z^k \rangle$$
$$+ \frac{2}{c}\langle \boldsymbol{\beta}^{k+1} - \boldsymbol{\beta}^*, \boldsymbol{\beta}^k - \boldsymbol{\beta}^{k+1} \rangle$$

Using definitions of $\mathbf{u}$ and $\mathbf{G}$, and doing some algebraic manipulations, we can rewrite inequality (13) as follows

$$m_f\|\mathbf{x}^{k+1} - \mathbf{x}^*\|_2^2 \leq \|\mathbf{u}^k - \mathbf{u}^*\|_G^2 - \|\mathbf{u}^{k+1} - \mathbf{u}^*\|_G^2$$
$$- \|\mathbf{u}^k - \mathbf{u}^{k+1}\|_G^2 + 2c\langle \mathbf{z}^{k+1} - \mathbf{z}^*, \mathbf{e}_z^k \rangle \tag{14}$$

In order to prove inequality (9), we need to show

$$2c\langle \mathbf{z}^{k+1} - \mathbf{z}^*, \mathbf{e}_z^k \rangle + \delta\|\mathbf{u}^{k+1} - \mathbf{u}^*\|_G^2$$
$$\leq m_f\|\mathbf{x}^{k+1} - \mathbf{x}^*\|_2^2 + \|\mathbf{u}^k - \mathbf{u}^{k+1}\|_G^2 \tag{15}$$

or equivalently

$$2c\langle \mathbf{z}^{k+1} - \mathbf{z}^*, \mathbf{e}_z^k \rangle + \delta c\|\mathbf{z}^{k+1} - \mathbf{z}^*\|_2^2 + \frac{\delta}{c}\|\boldsymbol{\beta}^{k+1} - \boldsymbol{\beta}^*\|_2^2$$
$$\leq m_f\|\mathbf{x}^{k+1} - \mathbf{x}^*\|_2^2 + c\|\mathbf{z}^k - \mathbf{z}^{k+1}\|_2^2 + \frac{1}{c}\|\boldsymbol{\beta}^k - \boldsymbol{\beta}^{k+1}\|_2^2 \tag{16}$$

We provide proper upper bounds on $\langle \mathbf{z}^{k+1} - \mathbf{z}^*, \mathbf{e}_z^k \rangle$, $\|\mathbf{z}^{k+1} - \mathbf{z}^*\|_2^2$ and $\|\boldsymbol{\beta}^{k+1} - \boldsymbol{\beta}^*\|_2^2$ below. First, we examine the first term in the LHS of (16). We have

$$2c\langle \mathbf{z}^{k+1} - \mathbf{z}^*, \mathbf{e}_z^k \rangle \leq 2c\|\mathbf{z}^{k+1} - \mathbf{z}^*\|_2 \|\mathbf{e}_z^k\|_2$$
$$\leq \frac{c}{2}\sigma_{max}(\mathbf{M}_+)\|\mathbf{x}^{k+1} - \mathbf{x}^*\|_2 \|\mathbf{e}_z^k\|_2 \quad (17)$$

Next, we bound the second term in (16). From (12c) we have

$$\|\mathbf{z}^{k+1} - \mathbf{z}^*\|_2^2 \leq \frac{1}{4}\sigma_{max}^2(\mathbf{M}_+)\|\mathbf{x}^{k+1} - \mathbf{x}^*\|_2^2 \quad (18)$$

About bounding the third term in (16), since $f(\mathbf{x})$ has Lipschitz continuous gradient with a constant $M_f$, we have $M_f^2\|\mathbf{x}^{k+1} - \mathbf{x}^*\|_2^2 \geq \|\nabla f(\mathbf{x}^{k+1}) - \nabla f(\mathbf{x}^*)\|_2^2$. In addition $\sigma_{max}^2(\mathbf{M}_+)\|\mathbf{z}^{k+1} - \mathbf{z}^k\|_2^2 \geq \|\mathbf{M}_+(\mathbf{z}^k - \mathbf{z}^{k+1})\|_2^2$. Using these two inequalities, and for any $\mu > 1$

$$c\sigma_{max}^2(\mathbf{M}_+)\|\mathbf{z}^{k+1} - \mathbf{z}^k\|_2^2 + (\mu-1)M_f^2\|\mathbf{x}^{k+1} - \mathbf{x}^*\|_2^2$$
$$\geq c\|\mathbf{M}_+(\mathbf{z}^k - \mathbf{z}^{k+1})\|_2^2 + (\mu-1)\|\nabla f(\mathbf{x}^{k+1}) - \nabla f(\mathbf{x})\|_2^2 \quad (19)$$

Accoring to (12a) and applying inequality $\|\mathbf{a} + \mathbf{b}\|_2^2 + (\mu-1)\|\mathbf{a}\|_2^2 \geq (1 - \frac{1}{\mu})\|\mathbf{b}\|_2^2$ to the RHS of (19) we obtain

$$c\sigma_{max}^2(\mathbf{M}_+)\|\mathbf{z}^{k+1} - \mathbf{z}^k\|_2^2 + (\mu-1)M_f^2\|\mathbf{x}^{k+1} - \mathbf{x}^*\|_2^2$$
$$\geq (1 - \frac{1}{\mu})\|\mathbf{M}_-(\boldsymbol{\beta}^{k+1} - \boldsymbol{\beta}^*) - c\mathbf{M}_+\mathbf{e}_z^k\|_2^2$$

Doing algebraic manipulations and taking into account that $\boldsymbol{\beta}^0$ lies in the column space of $\mathbf{M}_-^T$, we obtain

$$c\sigma_{max}^2(\mathbf{M}_+)\|\mathbf{z}^{k+1} - \mathbf{z}^k\|_2^2 + (\mu-1)M_f^2\|\mathbf{x}^{k+1} - \mathbf{x}^*\|_2^2$$
$$\geq (1 - \frac{1}{\mu})\{\tilde{\sigma}_{min}^2(\mathbf{M}_-)\|\boldsymbol{\beta}^{k+1} - \boldsymbol{\beta}^*\|_2^2$$
$$- 2c\sigma_{max}(\mathbf{M}_+)\sigma_{max}(\mathbf{M}_-)\|\boldsymbol{\beta}^{k+1} - \boldsymbol{\beta}^*\|_2 \|\mathbf{e}_z^k\|_2\}$$

which gives

$$\frac{1}{c}\|\boldsymbol{\beta}^{k+1} - \boldsymbol{\beta}^*\|_2^2 \leq \frac{2(\mu-1)M_f^2\|\mathbf{x}^{k+1} - \mathbf{x}^*\|_2^2}{c(1-\frac{1}{\mu})\tilde{\sigma}_{min}^2(\mathbf{M}_-)}$$
$$+ \frac{2c\sigma_{max}^2(\mathbf{M}_+)\|\mathbf{z}^{k+1} - \mathbf{z}^k\|_2^2}{(1-\frac{1}{\mu})\tilde{\sigma}_{min}^2(\mathbf{M}_-)} \quad (20)$$
$$+ \frac{4c\sigma_{max}^2(\mathbf{M}_+)\sigma_{max}^2(\mathbf{M}_-)\|\mathbf{e}_z^k\|_2^2}{\tilde{\sigma}_{min}^4(\mathbf{M}_-)}.$$

Combining (17), (18) and (20) we obtain

$$\frac{2c}{\delta}\langle \mathbf{z}^{k+1} - \mathbf{z}^*, \mathbf{e}_z^k \rangle + c\|\mathbf{z}^{k+1} - \mathbf{z}^*\|_2^2 + \frac{1}{c}\|\boldsymbol{\beta}^{k+1} - \boldsymbol{\beta}^*\|_2^2 \leq$$
$$\frac{c}{2\delta}\sigma_{max}(\mathbf{M}_+)\|\mathbf{x}^{k+1} - \mathbf{x}^*\|_2\|\mathbf{e}_z^k\|_2 + \frac{c}{4}\sigma_{max}^2(\mathbf{M}_+)\|\mathbf{x}^{k+1} - \mathbf{x}^*\|_2^2$$
$$+ \frac{2(\mu-1)M_f^2\|\mathbf{x}^{k+1} - \mathbf{x}^*\|_2^2}{c(1-\frac{1}{\mu})\tilde{\sigma}_{min}^2(\mathbf{M}_-)} + \frac{2c\sigma_{max}^2(\mathbf{M}_+)\|\mathbf{z}^{k+1} - \mathbf{z}^k\|_2^2}{(1-\frac{1}{\mu})\tilde{\sigma}_{min}^2(\mathbf{M}_-)} \quad (21)$$
$$+ \frac{4c\sigma_{max}^2(\mathbf{M}_+)\sigma_{max}^2(\mathbf{M}_-)\|\mathbf{e}_z^k\|_2^2}{\tilde{\sigma}_{min}^4(\mathbf{M}_-)}$$

Assume that $\|\mathbf{e}_z^k\|_2 \leq \|\mathbf{x}^{k+1} - \mathbf{x}^*\|_2$, hence we have

$$\frac{2c}{\delta}\langle \mathbf{z}^{k+1} - \mathbf{z}^*, \mathbf{e}_z^k \rangle + c\|\mathbf{z}^{k+1} - \mathbf{z}^*\|_2^2 + \frac{1}{c}\|\boldsymbol{\beta}^{k+1} - \boldsymbol{\beta}^*\|_2^2$$
$$\leq \{\frac{c}{2\delta}\sigma_{max}(\mathbf{M}_+) + \frac{c}{4}\sigma_{max}^2(\mathbf{M}_+) + \frac{2(\mu-1)M_f^2}{c(1-\frac{1}{\mu})\tilde{\sigma}_{min}^2(\mathbf{M}_-)}$$
$$+ \frac{4c\sigma_{max}^2(\mathbf{M}_+)\sigma_{max}^2(\mathbf{M}_-)}{\tilde{\sigma}_{min}^4(\mathbf{M}_-)}\}\|\mathbf{x}^{k+1} - \mathbf{x}^*\|_2^2$$
$$+ \frac{2c\sigma_{max}^2(\mathbf{M}_+)\|\mathbf{z}^{k+1} - \mathbf{z}^k\|_2^2}{(1-\frac{1}{\mu})\tilde{\sigma}_{min}^2(\mathbf{M}_-)}$$

Considering definition of $a$ and $b$ in Theorem 1, and according to inequality (16), we have

$$\delta = \min\left\{\frac{m_f}{\frac{c}{2\delta}\sigma_{max}(\mathbf{M}_+) + a}, \frac{1}{b}\right\} \quad (22)$$

$\delta = 0$ is a trivial solution of (22). If the first term is the minimizer of the RHS, then we have

$$\delta = \frac{m_f}{\frac{c}{2\delta}\sigma_{max}(\mathbf{M}_+) + a} \Rightarrow \delta = \frac{m_f - \frac{c\sigma_{max}(\mathbf{M}_+)}{2}}{a} \quad (23)$$

which implies that $m_f \geq \frac{c\sigma_{max}(\mathbf{M}_+)}{2}$ and subsequently $\delta = \min\left\{\frac{m_f - \frac{c\sigma_{max}(\mathbf{M}_+)}{2}}{a}, \frac{1}{b}\right\}$. Combining (14) and (18) and noting $\|\mathbf{e}_z^k\|_2 \leq \|\mathbf{x}^{k+1} - \mathbf{x}^*\|_2$, (10) can be proved. □

Suppose that condition $\|\mathbf{e}_z^k\|_2 \leq \|\mathbf{x}^{k+1} - \mathbf{x}^*\|_2$ is satisfied up to iteration $j$. Increase in $\|\mathbf{x}^k - \mathbf{x}^*\|_2$ for $k > j$ will eventually result in hold of $\|\mathbf{e}_z^k\|_2 \leq \|\mathbf{x}^{k+1} - \mathbf{x}^*\|_2$ condition which guarantees decrease in $\|\mathbf{u}^k - \mathbf{u}^*\|_G^2$.

**Corollary 1.** If $\|\mathbf{e}_x\|_2 = \sigma_e$ and constant over all iterations, an upper bound on the error $\|\mathbf{x}^k - \mathbf{x}^*\|_2$ for $k \to \infty$ is $\max_i(|\mathcal{N}_i|)\sigma_e$.

*Proof.* According to definition of $\mathbf{M}_+$, $\frac{1}{2}\mathbf{M}_+\mathbf{M}_+^T$ is equivalent to network signless Laplacian matrix $\mathbf{L}$. Consider the maximum eigenvalue of $\mathbf{L}$ as $l_{max}$. It can be shown that [18]

$$l_{max} \leq 2\max_i(|\mathcal{N}_i|) \quad (24)$$

Hence $\sigma_{max}(\mathbf{M}_+) \leq 2\sqrt{\max_i(|\mathcal{N}_i|)}$. Using definition of $\mathbf{e}_z$ and Theorem 1, we have

$$\|\mathbf{x}^k - \mathbf{x}^*\|_2 \leq \sqrt{\max_i(|\mathcal{N}_i|)}\sigma_e \quad k \to \infty \quad (25)$$

□

**Remark 1.** According to (25), a smaller $\max_i(|\mathcal{N}_i|)$ may improve the estimation accuracy, however, based on Theorem 1 it may in turn degrade the convergence rate.

**Remark 2.** One can examine that in the scenario without computing error, the convergence rate bound in [16] is tighter than that in Theorem 1. However, the bound in Theorem 1 is applicable to both scenarios with and without computing error, whereas the one in [16] only suites the setting without error.

## IV. NUMERICAL RESULTS

In this Section, we consider Algorithm 1 to solve the optimization problem $\min_{\tilde{\mathbf{x}}} \sum_{i=1}^{N} \frac{1}{2}\|\mathbf{y}_i - \mathbf{M}_i\tilde{\mathbf{x}}\|_2^2$. In this setup, $N = 200$ agents estimate variable $\tilde{\mathbf{x}}$ using noisy linear observations $\mathbf{y}_i = \mathbf{M}_i\tilde{\mathbf{x}} + \mathbf{n}_i$ cooperatively. The variables which are communicated between neighbors are subject to additive error as in (6); the error is assumed Gaussian $\mathcal{N}(0, \sigma_e^2)$. In our experiments, $\tilde{\mathbf{x}} \in \mathbb{R}^3$ and $\mathbf{M}_i \in \mathbb{R}^{3\times 3}$ and their elements are i.i.d from $\mathcal{N}(0,1)$. The observation noise is $\mathbf{n}_i \sim \mathcal{N}(\mathbf{0}, \sigma^2\mathbf{I}_3)$ and $\sigma^2 = 10^{-3}$. Define network connectivity ratio as $\rho = \frac{E}{E_c}$, where $E_c$ is the number of edges in a corresponding complete graph. Consider $\tilde{\mathbf{x}}_{i,k}^D$ and $\tilde{\mathbf{x}}^C$ as respectively the distributed and centralized estimates of $\tilde{\mathbf{x}}$ (at node $i$ and iteration $k$). Our performance metrics is error $\mathcal{E}_{i,k}^{DC} = \frac{\|\tilde{\mathbf{x}}_{i,k}^D - \tilde{\mathbf{x}}^C\|_2}{\|\tilde{\mathbf{x}}^C\|_2}$ or $E_i[\mathcal{E}_{i,k}^{DC}]$.

Figure 1 shows the average error of nodes at iteration $k$, $E_i[\mathcal{E}_{i,k}^{DC}]$, as a function of iteration number $k$ for different values of $\sigma_e$ and ADMM parameter $c$ computed over 100 random realizations of connected network $\mathcal{G}$ with a given connectivity ratio. In this experiment, the connectivity ratio is fixed and set to 0.04. As evident, the convergence rate is linear and the final value of estimation error is less than $\sigma_e$ (see Corollary 1). ADMM parameter $c$ play an important role in convergence rate of the algorithm. From this figure it is clear that change in $c$ parameter affects the convergence rate significantly and it seems important to set this value properly.

## V. CONCLUSIONS

We analyzed the convergence behavior of distributed ADMM for consensus optimization in presence of additive computation error, which is for example due to observation or quantization noise at the nodes. Specifically, we showed that (a noisy) ADMM converges linearly under certain conditions and also examined the associated convergence point. Numerical results in the case of collaborative mean squared error estimation was presented. Next steps of research include analytical assessment of optimized ADMM parameters in this setting and with different system and network parameters.

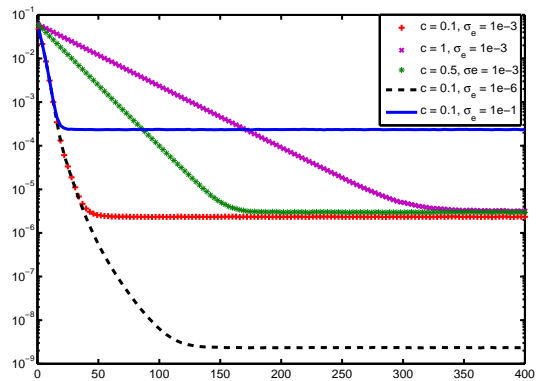

Fig. 1. Performance of noisy distributed ADMM, error reduction vs. iteration as a function of $\sigma_e$ and $c$. $\rho = 0.04$.

## REFERENCES

[1] S. H. Low and D. E. Lapsley, "Optimization flow control-I: basic algorithm and convergence," *IEEE/ACM Transactions on Networking*, vol. 7, no. 6, pp. 861–874, 1999.
[2] M. Rabbat and R. Nowak, "Distributed optimization in sensor networks," in *ACM Proceedings of the 3rd international symposium on Information processing in sensor networks*, 2004, pp. 20–27.
[3] A. Nedić and A. Ozdaglar, "Distributed subgradient methods for multi-agent optimization," *IEEE Transactions on Automatic Control*, vol. 54, no. 1, pp. 48–61, 2009.
[4] B. Johansson, M. Rabi, and M. Johansson, "A randomized incremental subgradient method for distributed optimization in networked systems," *SIAM Journal on Optimization*, vol. 20, no. 3, pp. 1157–1170, 2009.
[5] E. Ghadimi, I. Shames, and M. Johansson, "Multi-step gradient methods for networked optimization," *IEEE Transactions on Signal Processing*, vol. 61, no. 21, pp. 5417–5429, 2013.
[6] N. Li, "Distributed optimization in power networks and general multi-agent systems," Ph.D. dissertation, California Institute of Technology, 2013.
[7] T.-H. Chang, A. Nedic, and A. Scaglione, "Distributed constrained optimization by consensus-based primal-dual perturbation method," *IEEE Transactions on Automatic Control*, vol. 59, no. 6, pp. 1524–1538, 2014.
[8] S. Boyd, N. Parikh, E. Chu, B. Peleato, and J. Eckstein, "Distributed optimization and statistical learning via the alternating direction method of multipliers," *Foundations and Trends in Machine Learning*, vol. 3, no. 1, pp. 1–122, 2011.
[9] W. Deng and W. Yin, "On the global and linear convergence of the generalized alternating direction method of multipliers," *Journal of Scientific Computing*, pp. 1–28, 2012.
[10] B. He and X. Yuan, "On the o(1/n) convergence rate of the douglas-rachford alternating direction method," *SIAM Journal on Numerical Analysis*, vol. 50, no. 2, pp. 700–709, 2012.
[11] T. Goldstein, B. O'Donoghue, S. Setzer, and R. Baraniuk, "Fast alternating direction optimization methods," *SIAM Journal on Imaging Sciences*, vol. 7, no. 3, pp. 1588–1623, 2014.
[12] L. Xiao, S. Boyd, and S.-J. Kim, "Distributed average consensus with least-mean-square deviation," *Journal of Parallel and Distributed Computing*, vol. 67, no. 1, pp. 33–46, 2007.
[13] B. Johansson, T. Keviczky, M. Johansson, and K. H. Johansson, "Subgradient methods and consensus algorithms for solving convex optimization problems," in *47th IEEE Conference on Decision and Control*, 2008, pp. 4185–4190.
[14] G. Mateos, J. A. Bazerque, and G. B. Giannakis, "Distributed sparse linear regression," *IEEE Transactions on Signal Processing*, vol. 58, no. 10, pp. 5262–5276, 2010.
[15] J. F. Mota, J. M. Xavier, P. M. Aguiar, and M. Puschel, "D-admm: A communication-efficient distributed algorithm for separable optimization," *IEEE Transactions on Signal Processing*, vol. 61, no. 10, pp. 2718–2723, 2013.
[16] W. Shi, Q. Ling, K. Yuan, G. Wu, and W. Yin, "On the linear convergence of the admm in decentralized consensus optimization," *IEEE Transactions on Signal Processing*, vol. 62, no. 7, pp. 1750–1761, 2014.
[17] T.-H. Chang, M. Hong, and X. Wang, "Multi-agent distributed optimization via inexact consensus admm," *IEEE Transactions on Signal Processing*, vol. 63, no. 2, pp. 482–497, 2015.
[18] D. Cvetković, P. Rowlinson, and S. K. Simić, "Signless laplacians of finite graphs," *Linear Algebra and its applications*, vol. 423, no. 1, pp. 155–171, 2007.